\documentclass[12pt,preprint]{aastex}
\usepackage{graphicx}

\begin{document}

\title{Molecular Line Observations of a Carbon-Chain-Rich Core L492}
\author{Tomoya HIROTA}
\affil {National Astronomical Observatory of Japan,}
\affil {Osawa 2-21-1, Mitaka, Tokyo 181-8588, Japan; 
tomoya.hirota@nao.ac.jp}
\and
\author{Satoshi YAMAMOTO}
\affil {Department of Physics and Research Center for the Early Universe,}
\affil {The University of Tokyo, Bunkyo-ku, Tokyo 113-0033, JAPAN }

\begin{abstract}

We report on molecular abundances and distributions in a starless dense core L492. 
We have found that the abundances of carbon-chain molecules such as 
CCS, C$_{3}$S, HC$_{3}$N, HC$_{5}$N, and HC$_{7}$N are comparable to those in 
chemically young dark cloud cores called "carbon-chain--producing regions", 
such as L1495B, L1521B, L1521E, and TMC-1. 
This is the first dark cloud core with extremely rich in carbon-chain-molecules 
that is found outside the Taurus region. 
In addition, the deuterium fractionation ratios of 
DNC/HNC and DCO$^{+}$/HCO$^{+}$ are also comparable to those in 
carbon-chain--producing regions, 
being significantly lower than those in 
the evolved prestellar cores such as L1498 and L1544. 
On the other hand, the abundances of NH$_{3}$ and N$_{2}$H$^{+}$ 
are systematically higher than those in carbon-chain--producing regions. 
Our mapping observations reveal that the central hole of molecular distributions, 
which were reported for CCS and C$^{34}$S in 
evolved prestellar cores is not significant in L492, 
indicating that the depletion factor of molecules is not very high. 
Furthermore, L492 is dynamically more evolved than carbon-chain--producing regions, 
and the protostellar collapse has started like L1498 and L1544.
Therefore, it is likely that the chemical and dynamical evolutionary stage of 
L492 is intermediate between 
carbon-chain--producing regions (L1495B, L1521B, L1521E, and TMC-1) 
and evolved prestellar cores (L1498 and L1544). 
\end{abstract}


\section{Introduction}

In the last decade, special attention has been paid to 
starless cores or prestellar cores in dark clouds 
in order to investigate initial conditions of protostellar collapse. 
Since the detection of infalling motion in a prestellar core L1544 in 
the Taurus Molecular Cloud \citep{mye96}, 
a number of observational studies on L1544 and similar prestellar cores 
such as L1498 in the Taurus Molecular Cloud have been carried out 
with molecular lines and dust continuum emission \citep[e.g.][]{taf02, taf04b}. 
One of the most interesting features found in L1498 and L1544 
is an evidence of molecular depletion in the central part of the cores 
\citep[e.g.][]{kui96, oha99, cas02b}. 
Both observational and theoretical studies reveal that 
abundances of CO, CS, and CCS are depleted in the chemically evolved cores, 
while NH$_{3}$ and N$_{2}$H$^{+}$ are not \citep{ber97, aik05}. 
These studies have promoted understanding of 
chemical properties as well as kinematics of evolved prestellar cores 
just under the collapsing phase. 

Although detailed studies on evolved prestellar cores were extensively carried out, 
no systematic studies on younger cores have been made for a long time, 
except for TMC-1 \citep[e.g.][]{hir92, hir95, pra97}. 
It is because most of dense cores observed in the previous studies 
were based on the catalog of \citet{ben89}, which are prepared by the NH$_{3}$ observations. 
On the other hand, \citet{suz92} pointed out that NH$_{3}$ is not always 
a good tracer of dense cores because of the chemical abundance 
variation from core to core. They found anticorrelation between abundances of 
NH$_{3}$ and carbon-chain-molecules such as CCS, HC$_{3}$N, and HC$_{5}$N, 
and identified a few starless cores where the carbon-chain molecules are abundant 
while NH$_{3}$ is deficient. 
They are L1495B, L1521B, L1521E, and the cyanopolyyne peak of TMC-1, and are 
called "carbon-chain--producing regions". 
Chemical model calculations suggest that the carbon-chain--producing regions 
are in the early stage of chemical evolution while dense cores traced 
by the NH$_{3}$ lines are rather evolved \citep{suz92, ber97}. 

In order to investigate basic physical and chemical properties of carbon-chain--producing 
regions, we carried out detailed observations of L1521E \citep{hir02}, 
L1495B and L1521B \citep{hir04}. 
For each of these three regions, 
we found a compact dense core traced by the H$^{13}$CO$^{+}$, 
CCS, C$_{3}$S, and HC$_{3}$N lines and their distributions have 
a single peak at the same position. 
Such a distribution is different from those in evolved 
prestellar cores, L1498 and L1544. 
In addition, abundances of NH$_{3}$ and N$_{2}$H$^{+}$ are extremely lower 
in these three regions than in other prestellar cores, 
while those of carbon-chain-molecules are systematically higher. 
According to our survey of deuterated molecules, deuterium fractionation ratios of 
DNC/HNC and DCO$^{+}$/HCO$^{+}$ are systematically lower in L1495B, L1521B, and 
L1521E than in the other dark cores \citep{hir01, hir03}. 
These features suggest that L1495B, L1521B, and L1521E would be in 
the early stage of dynamical and chemical evolution, and the degree of depletion 
of molecules is possibly lower than in other evolved cores \citep{taf04}. 

Although detailed studies on carbon-chain--producing regions are important for understanding 
of chemical and physical evolution of dense cores, only four carbon-chain--producing regions 
were identified in Taurus Molecular Cloud. In order to search for 
carbon-chain--producing regions outside the Taurus Molecular Cloud, we carried out 
a survey of CCS, HC$_{3}$N, and HC$_{5}$N toward 31 dark cloud cores. As a result, 
we detected a possible candidate for another carbon-chain--producing region L492 in 
the Aquila Rift (Hirota et al. in preparation). 
L492 is one of the prestellar cores 
identified as a strong infalling candidate \citep{lee01}, and 
it is observed with the dust continuum emission as well as the 
N$_{2}$D$^{+}$ and N$_{2}$H$^{+}$ lines \citep{cra05}.  
In this paper, we report on the results of extensive molecular line observations of L492. 

\section{Observations}

We carried out molecular line observations of L492 with the 45 m radio 
telescope at Nobeyama Radio Observatory (NRO)\footnotemark 
\footnotetext{Nobeyama Radio Observatory is a branch of 
the National Astronomical Observatory of Japan, an interuniversity 
research institute operated by the Ministry of 
Education, Science, Sports and Culture of Japan} 
in February 2003 and March 2004. 
The observed lines are summarized in Table \ref{tab-observe}. 

We used cooled HEMT receivers for the lines in the 22-23 GHz band, 
SIS mixer receivers for the lines in the 45-48 GHz and 72-96 GHz bands, 
the SIS 25-beam array receiver system (BEARS) for the 
$^{13}$CO and C$^{18}$O lines in the 109-110 GHz band. 
The main-beam efficiencies ($\eta_{mb}$) were 0.8, 0.7, 0.5, and 
0.5 for the 22-23 GHz, 45-49 GHz, 72-96 GHz, and 109-110 GHz 
bands, respectively. 
The beam sizes were 73\arcsec, 36\arcsec, 20\arcsec, and 16\arcsec \ 
for the 22-23 GHz, 45-49 GHz, 72-96 GHz, and 109-110 GHz 
bands, respectively. 
Acousto-optical radio spectrometers with the frequency resolution of 
37 kHz were used for the backend except for the $^{13}$CO and 
C$^{18}$O observations, in which autocorrelators with the frequency 
resolution of 31.25 kHz were used. 

\begin{figure}
\epsscale{1}
\plotone{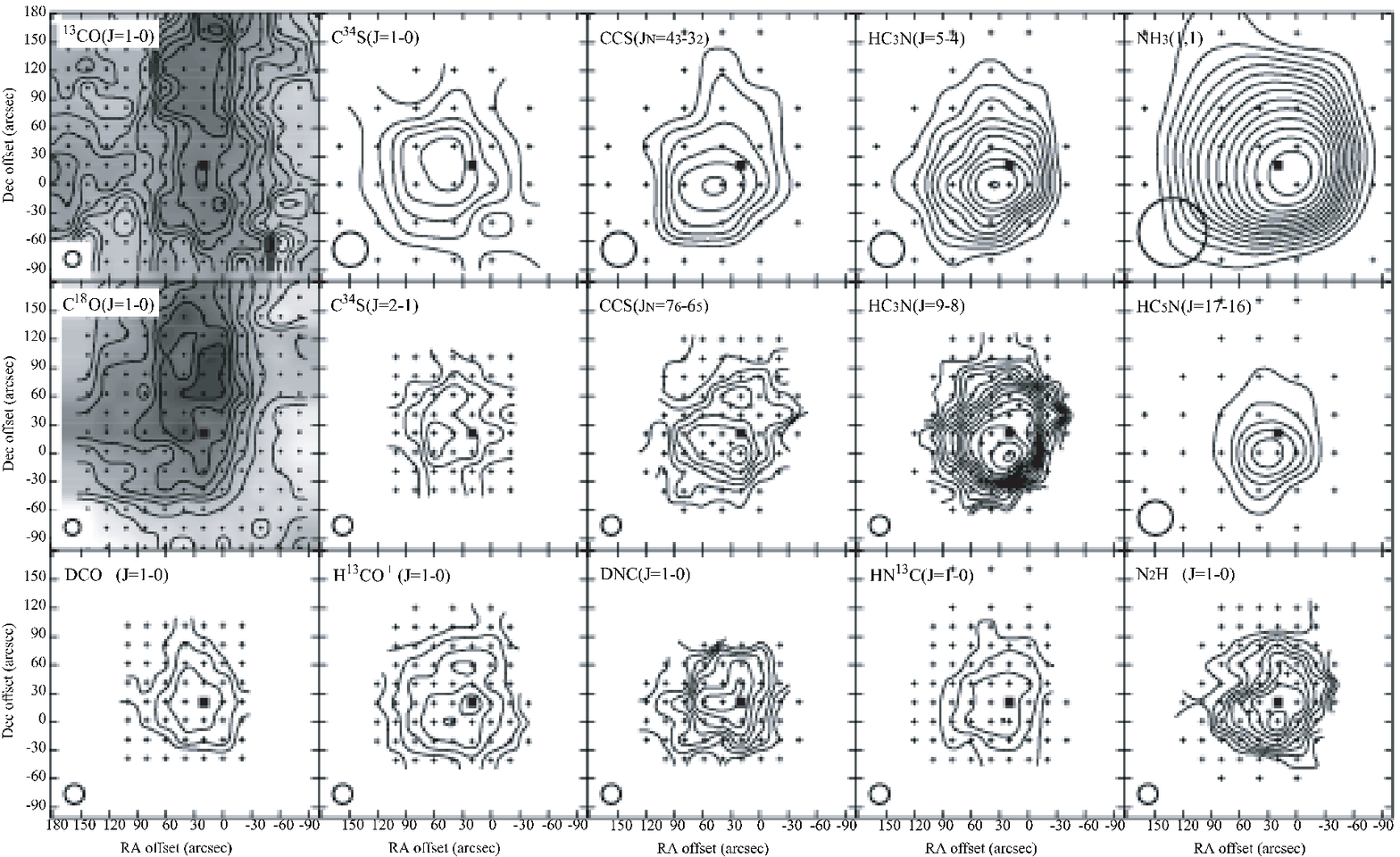}
\caption{Integrated intensity maps of molecular lines in L492.
The velocity range of integration (in km s$^{-1}$), 
the interval of the contours ($2\sigma$, in K km s$^{-1}$) and 
the lowest contour ($3\sigma$, in K km s$^{-1}$) are 
(7.0-9.0, 0.24, 0.36) for $^{13}$CO($J$=1-0), 
(7.0-8.5, 0.070, 0.105) for C$^{34}$S($J$=1-0), 
(7.2-8.3, 0.100, 0.150) for CCS($4_{3}$-$3_{2}$), 
(7.0-8.5, 0.200, 0.300) for HC$_{3}$N($J$=5-4), 
(7.2-8.6, 0.100, 0.150) for NH$_{3}$(1,1), 
(7.2-8.8, 0.150, 0.225) for C$^{18}$O($J$=1-0), 
(7.3-8.2, 0.050, 0.075) for C$^{34}$S($J$=2-1), 
(7.2-8.3, 0.060, 0.090) for CCS($7_{6}$-$6_{5}$), 
(7.2-8.3, 0.060, 0.090) for HC$_{3}$N($J$=9-8), 
(7.3-8.3, 0.100, 0.150) for HC$_{5}$N($J$=17-16), 
(7.2-8.3, 0.110, 0.165) for DCO$^{+}$($J$=1-0), 
(7.2-8.5, 0.080, 0.120) for H$^{13}$CO$^{+}$($J$=1-0), 
(7.0-8.5, 0.070, 0.105) for DNC($J$=1-0), 
(7.0-8.5, 0.080, 0.120) for HN$^{13}$C($J$=1-0), and 
($-0.6$-15.2, 0.200, 0.300) for N$_{2}$H$^{+}$($J$=1-0, total). 
For the NH$_{3}$(1,1) line, the integrated intensity is the sum of all 5 
components (see text and Table \ref{tab-observe}). 
For the N$_{2}$H$^{+}$($J$=1-0) line, the velocity range of integration is 
for the $F_{1},F$=2,3-1,2 component. 
The beam size (HPBW) is shown in the lower left of each map. 
The peak position of the dust continuum emission \citep{cra05} is 
indicated by a filled square, and observed positions are indicated by 
small crosses. 
Gray scale maps are superposed to the $^{13}$CO($J$=1-0) and C$^{18}$O($J$=1-0) 
maps to clarify their complex structures. 
\label{fig-maps}}
\end{figure}

We took the reference position of L492 to be 
$\alpha_{2000}=18^{h}15^{m}46^{s}.1$, 
$\delta_{2000}=-03^{\circ}$46\arcmin13\arcsec \citep{lee01, cra05}. 
Observations in the 22-23 GHz and 45-49 GHz bands 
were performed in the position-switching mode. 
The off position was taken to be 10\arcmin \ away from the source position, 
which is outside the core of L492 with the radius of 3.2\arcmin \citep{lee99a}. 
Observations in the 72-96 GHz and 109-110 GHz bands were performed 
in the frequency switching mode with the frequency offset was set to be 
7.685 MHz. 
The grid spacing in the mapping observations was taken to be 
40-80\arcsec, 40\arcsec, and 20\arcsec \ for the 22-23 GHz, 45-49 GHz, 
and 72-96 GHz bands, respectively, as shown in Figure \ref{fig-maps}. 
Since the beam separation of 
the BEARS is 41.1\arcsec, the grid spacing of the $^{13}$CO and 
C$^{18}$O maps were taken to be 20.55\arcsec. Note that 
we did not carry out the Niquist sampling because of limited observing 
time. Pointing was checked by observing a nearby SiO maser source, 
R-Aql, every 1-2 hours, and the pointing accuracy was 
estimated to be better than 5\arcsec. 
The antenna temperature was calibrated by a usual chopper-wheel method. 

\section{Results}

\subsection{Molecular Distributions and Velocity Structure} 

\begin{figure}
\epsscale{1}
\plotone{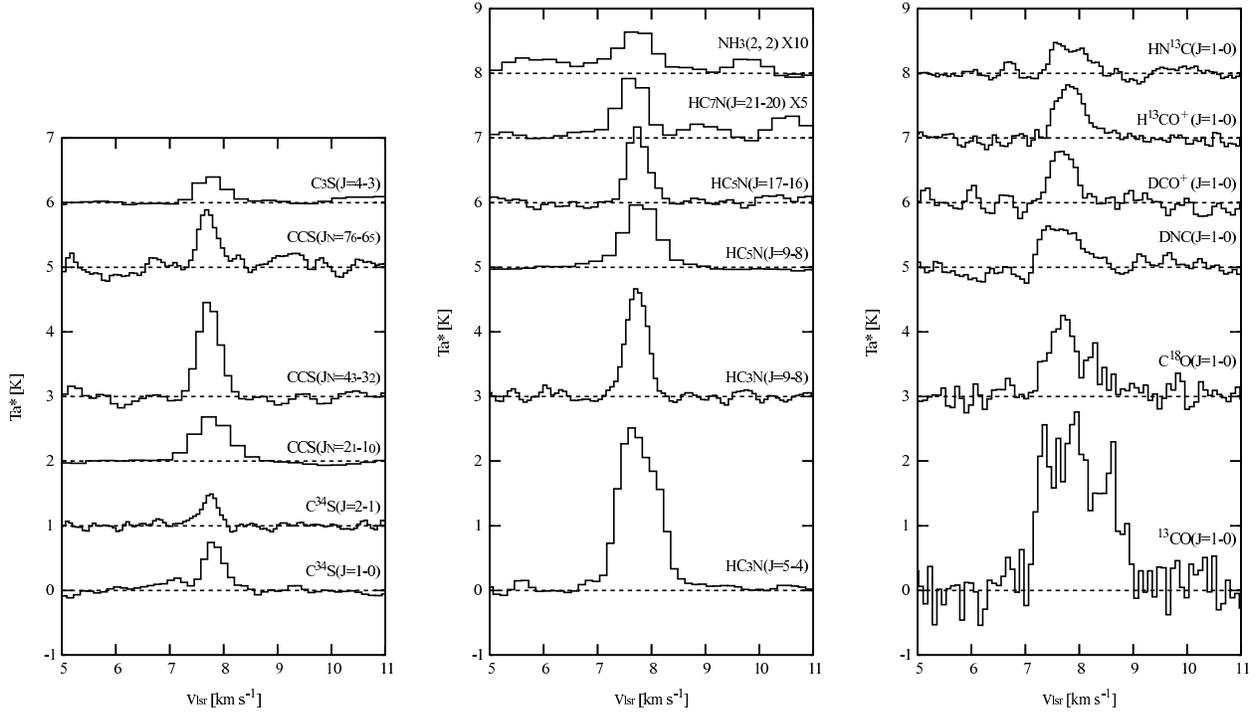}
\caption{Examples of the observed spectral lines. 
{\it{Left}}: Spectra of sulpher-bearing molecules. They are observed toward the 
(40\arcsec, 0\arcsec) position of L492. 
{\it{Middle}}: Spectra of cyanopolyynes and NH$_{3}$. 
They are observed toward the (40\arcsec, 0\arcsec) position of L492, 
except for NH$_{3}$(2,2) and HC$_{7}$N($J$=21-20), which are 
observed at the (0\arcsec, 0\arcsec) position. 
{\it{Right}}: Spectra of deuterated molecules and CO isotopes.
They are observed toward the (20\arcsec, 20\arcsec) position of L492. 
\label{fig-spectr}}
\end{figure}

\begin{figure}
\epsscale{0.7}
\plotone{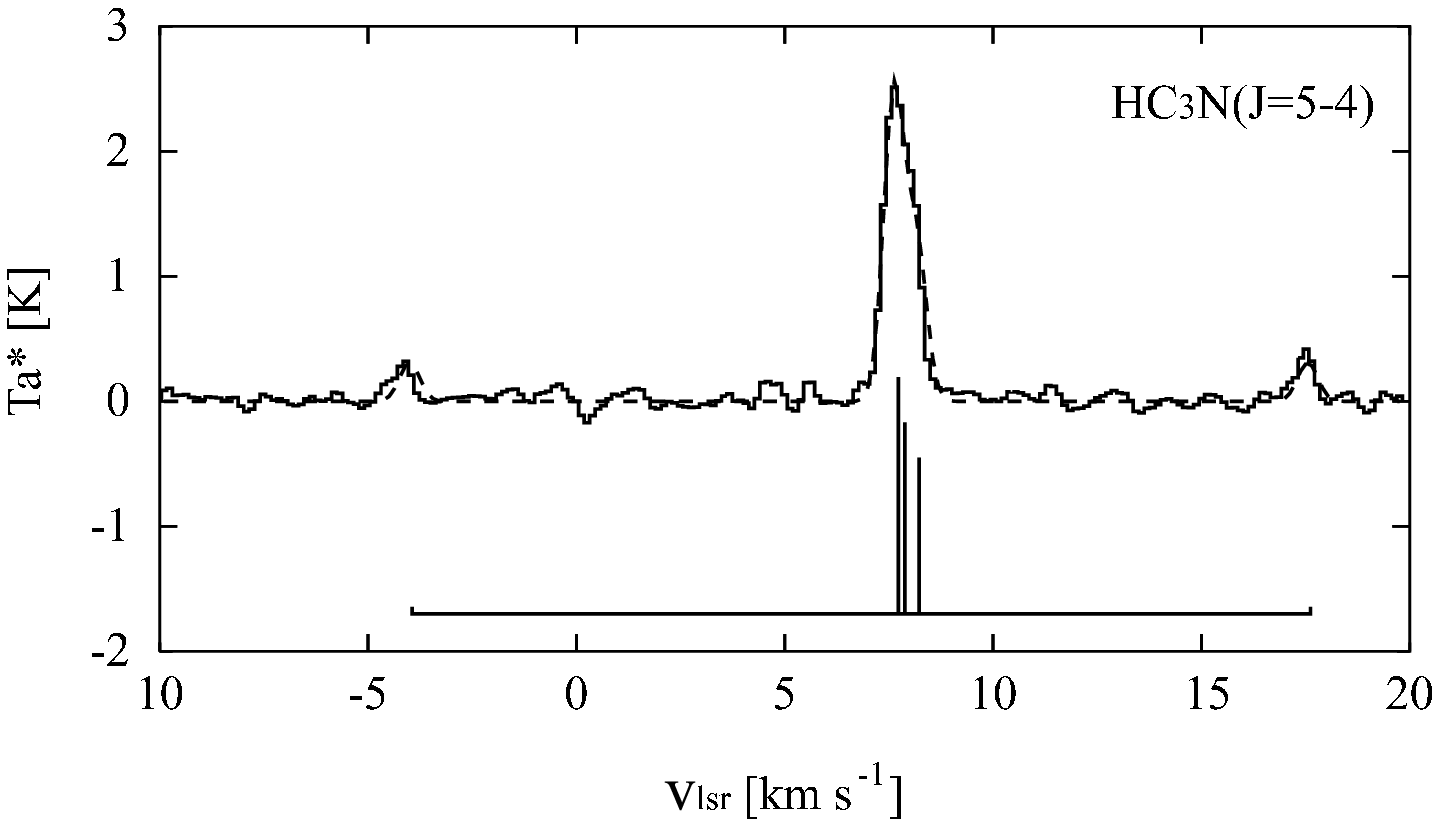}
\caption{A spectrum of HC$_{3}$N($J$=5-4) at the (40\arcsec, 0\arcsec) position of L492. 
A dashed line represents the best fit model. 
The expected hyperfine pattern of the HC$_{3}$N 
line is indicated in the bottom of the figure. 
\label{fig-sphc3n} }
\end{figure}

\begin{figure}
\epsscale{0.7}
\plotone{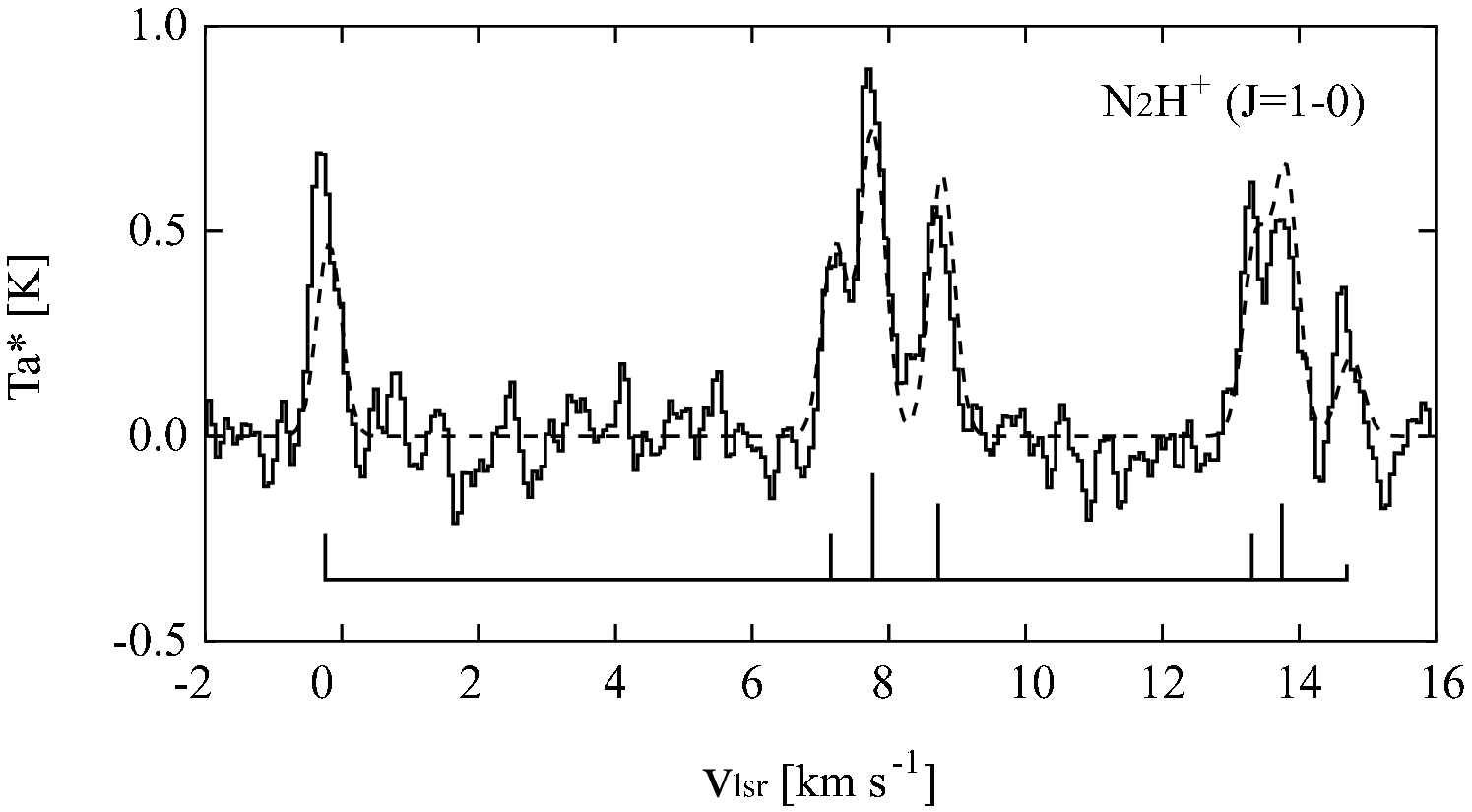}
\caption{A spectrum of N$_{2}$H$^{+}$($J$=1-0) 
at the (20\arcsec, 20\arcsec) position of L492. 
A dashed line represents the best fit model. 
The expected hyperfine pattern of the N$_{2}$H$^{+}$ 
line is indicated in the bottom of the figure. 
\label{fig-spn2hp} }
\end{figure}

\begin{figure}
\epsscale{0.7}
\plotone{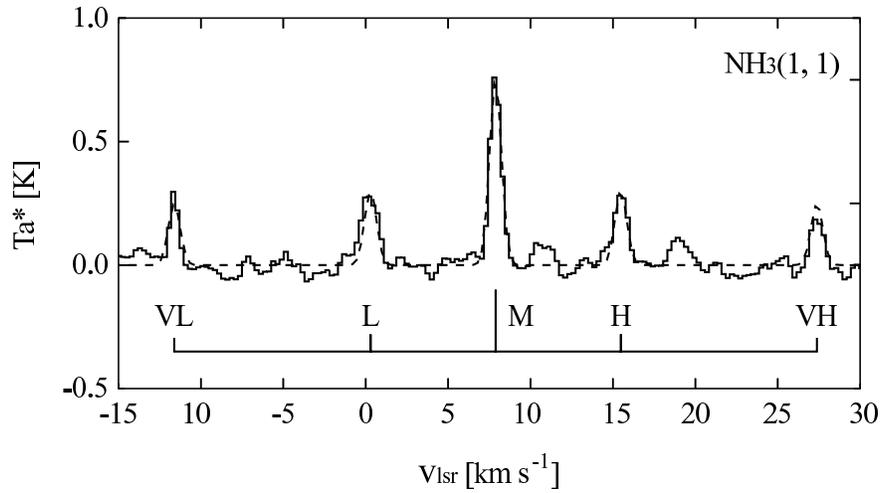}
\caption{A spectrum of NH$_{3}$($J,K$=1,1) 
at the (0\arcsec, 0\arcsec) position of L492. 
A dashed line represents the best fit model. 
The expected hyperfine pattern of the NH$_{3}$ 
line are indicated in the bottom of the figure. 
Groups of blended hyperfine components are labeled as adopted by 
\citet{ung80}. 
\label{fig-spnh3} }
\end{figure}

Figure \ref{fig-maps} shows the integrated intensity maps, 
and Figures \ref{fig-spectr}-\ref{fig-spnh3} show an example of 
the observed spectrum for each molecule. 
We only show the maps of the lines marked in Table \ref{tab-observe}. 
%
%
The molecular gas is distributed along the north-south ridge traced by 
the $^{13}$CO and C$^{18}$O lines, and the dense core of L492 traced 
by other lines is located at the southern edge of the ridge. 
Most of the maps show quite similar structure with a single peak 
and a simple elliptical shape with the average diameter of 100\arcsec, 
corresponding to 0.1 pc at the distance of 200 pc \citep{lee99a}. 
Although the peak positions are different from each other, this is partly 
due to the different beam size and grid spacing. 
The single peak distribution of the C$^{34}$S and CCS in L492 
clearly contrasts with the double peak distribution in 
L1498 and L1544 \citep{kui96, oha99, taf02, taf04b}, and is rather 
similar to the distribution in the carbon-chain--producing regions 
such as L1495B, L1521B, and L1521E \citep{hir02, hir04}. 
The peak position of the N$_{2}$H$^{+}$ map is 20\arcsec \ south of 
those of the dust continuum, N$_{2}$D$^{+}$, and N$_{2}$H$^{+}$ maps reported 
by \citet{cra05}, probably due to insufficient spatial 
resolution or coarse grid spacing. 

On the other hand, we found a marginal double peak structure in the maps 
of the $J$=1-0 line of H$^{13}$CO$^{+}$ and the $7_{6}$-$6_{5}$ line of CCS, 
although the northern peak has the statistical significance of 
only 2 $\sigma$ level. 
Because the optical depth of the $J$=1-0 line of H$^{13}$CO$^{+}$ 
and the $7_{6}$-$6_{5}$ line of CCS are 0.82 and 
0.54, respectively, as discussed later, they would not be suffered from 
self-absorption. 
A possible explanation is depletion of H$^{13}$CO$^{+}$ and CCS 
in the central part of the core. 
Note that the double peak structure is not evident in the CCS 
$4_{3}$-$3_{2}$ map as mentioned above. 
Since the critical density for the $7_{6}$-$6_{5}$ line is higher 
than that of the $4_{3}$-$3_{2}$ line, we could detect the 
depletion of CCS only with the $7_{6}$-$6_{5}$ line. 
If the double peak structure is real, 
the depletion of molecules would have just started 
at the central part of L492, which can be seen only in 
the $7_{6}$-$6_{5}$ line of CCS and the $J$=1-0 line of H$^{13}$CO$^{+}$. 
Further high resolution observations would be necessary to confirm this. 

\begin{figure}
\epsscale{0.5}
\plotone{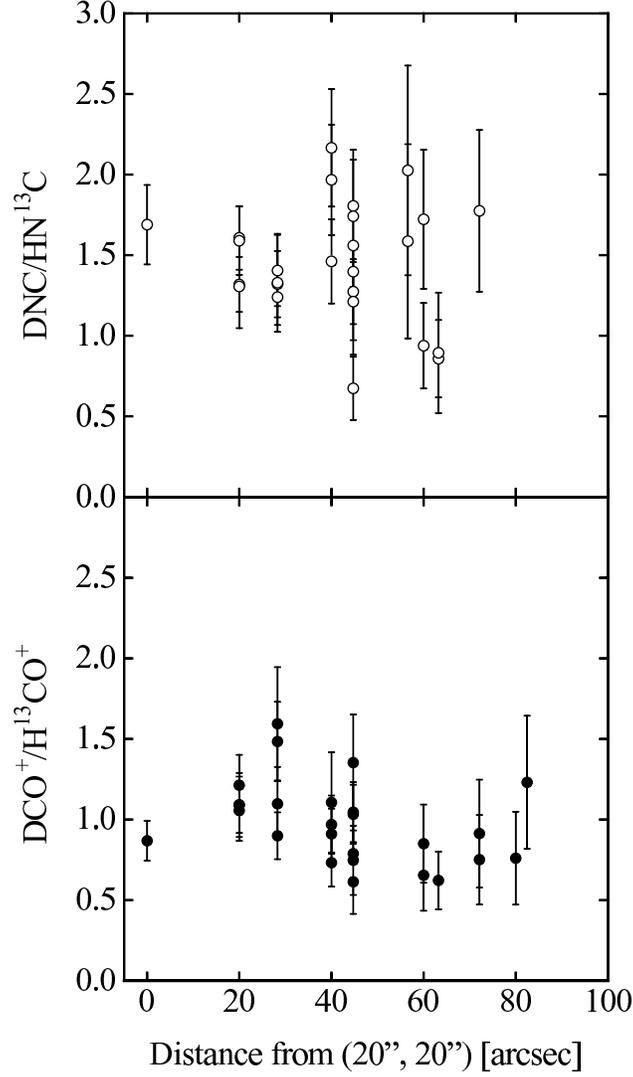}
\caption{Integrated intensity ratio of deuterated molecules 
as a function of the distance from the dust continuum peak. 
{\it{Top}}: DNC/HN$^{13}$C ratio. 
{\it{Bottom}}: DCO$^{+}$/H$^{13}$CO$^{+}$ ratio. 
Errors are estimated from the rms noise of each spectrum (2$\sigma$). 
Because these lines are optically thin, the integrated intensity ratios 
approximately represent the deuterium fractionation ratios as a function 
of the distance from the dust continuum peak (see text). 
\label{fig-dhratio}}
\end{figure}

It is interesting to compare the deuterium fractionation ratio within the dense core, 
because the deuterium fractionation ratio reflects chemical evolutionary stage and 
degree of depletion \citep{cas02b, hir01, hir03}. 
Figure \ref{fig-dhratio} shows 
the integrated intensity ratios of DCO$^{+}$/H$^{13}$CO$^{+}$ and DNC/HN$^{13}$C 
as a function of the distance from the dust continuum peak. 
Because these lines are optically thin as discussed later, the integrated 
intensity ratios approximately represent the abundance ratios. 
The maximum DCO$^{+}$/H$^{13}$CO$^{+}$ and DNC/HN$^{13}$C ratios are 
1.6 and 2.2, respectively, and their positions are offset 
from the DCO$^{+}$ and DNC peaks and also from the dust continuum peak. 
The deuterium fractionation ratios are not significantly enhanced in 
the central part of the core. We might be able to see an only slight 
increase in the DCO$^{+}$/H$^{13}$CO$^{+}$ ratio within the inner 
40\arcsec \ area. However, this is not so clear as in the case of 
L1544 \citep{cas02b}. As for the DNC/HN$^{13}$C ratio, 
such an enhancement is not seen, being consistent with other cores 
\citep{hir03}. 

\begin{figure}
\epsscale{0.7}
\plotone{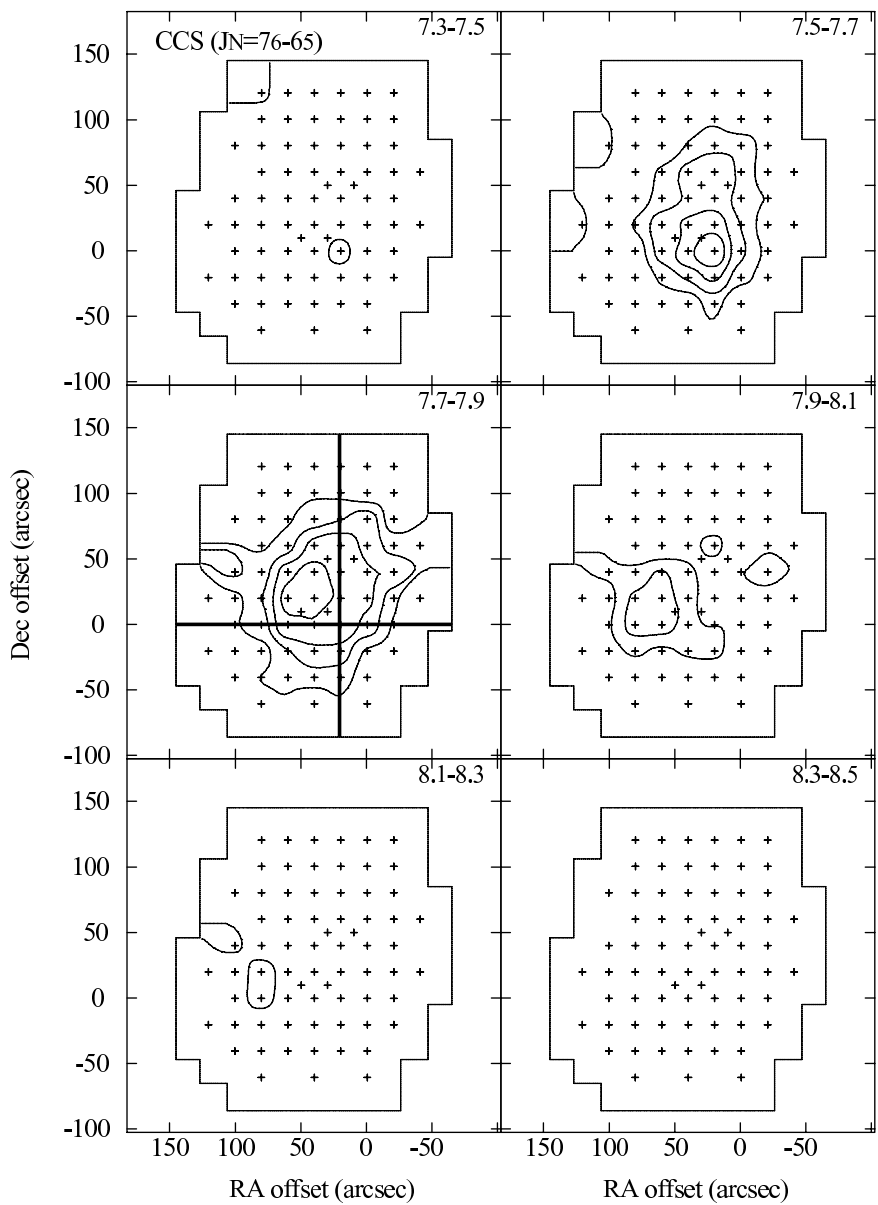}
\caption{Channel maps of the CCS($7_{6}-6_{5}$) line in L492.
The velocity range is labeled at the top right corner in each panel. 
The interval of the contours and the lowest contour are 0.030 K km s$^{-1}$ 
and 0.045 K km s$^{-1}$, respectively.  
The peak LSR velocity of L492 corresponds to 7.7-7.9 km s$^{-1}$. 
The position-velocity diagrams shown in Figure \ref{fig-l492-pv} are 
made along the bold solid lines in this panel. 
\label{fig-ccs81chmap2}}
\end{figure}

\begin{figure}
\epsscale{0.6}
\plotone{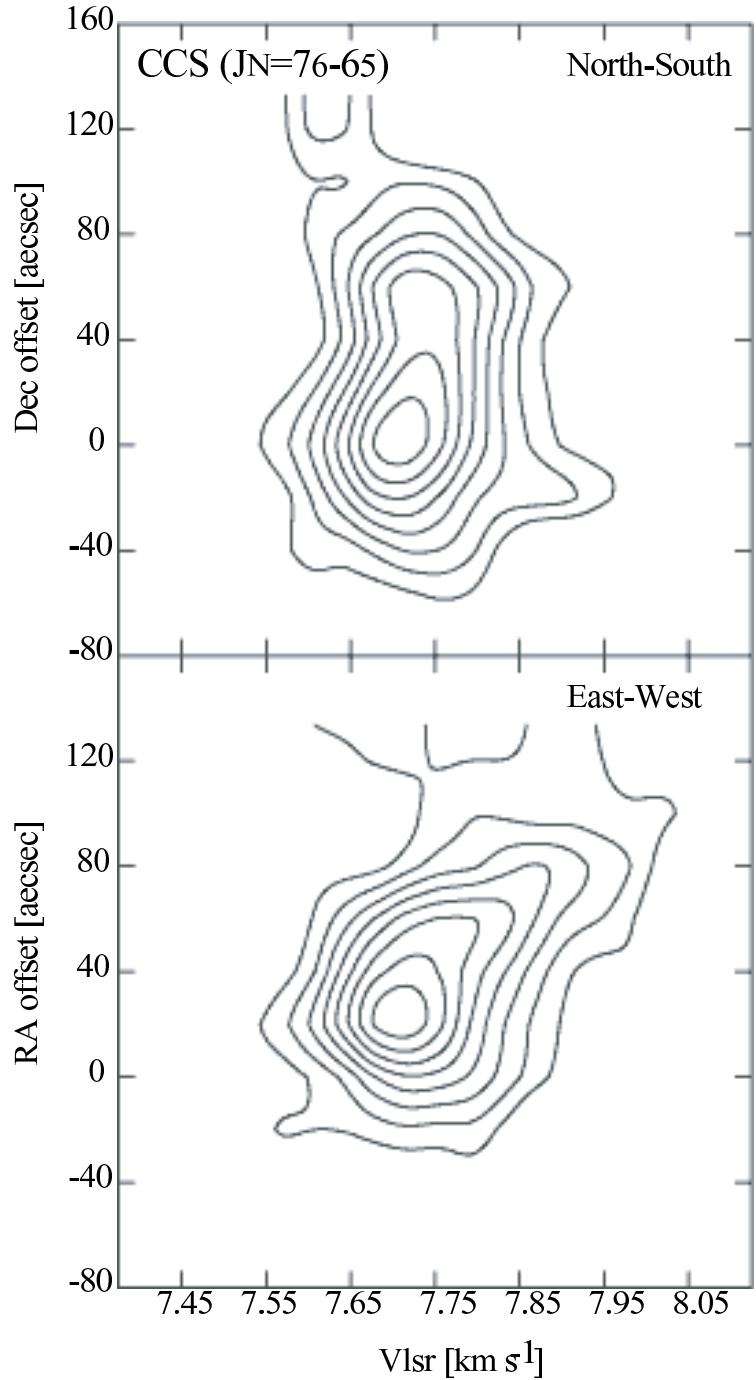}
\caption{Position-velocity (p-v) diagrams of the CCS($J_{N}$=$7_{6}$-$6_{5}$) line. 
The top panel shows 
the p-v diagram along the north-south direction at the offset in right ascension of 
20\arcsec, and the bottom panel shows that along the east-west direction 
at the offset in declination of 0\arcsec, 
as shown in Figure \ref{fig-ccs81chmap2}.
The interval of the contours and the lowest contour are 0.10 K 
for both panels. 
\label{fig-l492-pv}}
\end{figure}

We also investigate velocity structure of the core using the 
CCS ($7_{6}$-$6_{5}$) line, because this line has no hyperfine splitting 
and was observed with a relatively high velocity resolution (0.136 km s$^{-1}$). 
Figures \ref{fig-ccs81chmap2} and \ref{fig-l492-pv} show 
the channel maps and position-velocity diagrams of the CCS($7_{6}$-$6_{5}$) line, 
respectively. 
We found a global velocity gradient of 3 km s$^{-1}$ pc$^{-1}$ 
in the east-west direction across the core. 
On the other hand, we could not find other significant change in 
the linewidth \citep[e.g.][]{cra05}. 
According to \citet{lee01}, the optically thick line, CS($J$=2-1), toward L492 
has the asymmetric line profile due to an infalling motion. 
However, either double peak or asymmetric line profiles 
are not detected in our observations because we observed only optically thin lines. 

\subsection{Abundances for Individual Molecules}

In order to evaluate abundances of the observed molecules, 
line parameters for all the observed lines were determined by the 
Gaussian fit, as summarized in Table \ref{tab-observe}. 
We ignored unresolved hyperfine 
structures of the NH$_{3}$ and HC$_{3}$N lines, and hence, 
their linewidths are shown to be broader than the others. 
The parameters at the (0\arcsec, 0\arcsec) position for NH$_{3}$ and HC$_{7}$N 
(22-23 GHz band), the (40\arcsec,0\arcsec) position for other carbon-chain molecules 
(22-23 GHz, 45-49 GHz and 81-96 GHz bands), and 
the (20\arcsec, 20\arcsec) position for the rest (72-93 GHz and 109-110 GHz bands) 
are listed in Table \ref{tab-observe}. 
These positions correspond to the peak positions of the 
integrated intensity map for each molecule. 
Using these line parameters, 
we calculated the column densities of the observed molecules by the consistent way 
employed in previous works (e.g. Suzuki et al. 1992; Hirota et al. 1998, 2001, 2002, 2004) 
in order to compare the present results with those for L1495B, L1521B, L1521E, and TMC-1. 
The methods are briefly described below, and the results are 
summarized in Tables \ref{tab-lvg} and \ref{tab-lte}. 
The references for collision rates and dipole moments are also 
summarized in Tables \ref{tab-lvg} and \ref{tab-lte}. 
In the analysis, the kinetic temperature in L492 is assumed to be 10 K, 
which was derived from the NH$_{3}$ lines in the present study. 
Because we observed molecular lines with the different grid spacing
depending on the beam size, we can not compare the line parameters 
and column densities of all the observed molecules at the same position. 
Therefore, we list the line parameters and molecular column densities 
at the above positions. 
Although the peak positions are different from each other, this is partly due to the 
different beam size and grid spacing. Therefore, the column densities for the positions 
listed in Tables \ref{tab-lvg} and \ref{tab-lte} would be those for the same volume of gas. 
According to our maps shown in Figure \ref{fig-maps}, the difference in the 
integrated intensities due to the position difference is less than 20\%, 
and hence, the uncertainty in the column densities due to the 
position difference is estimated to be 20\%. 

\subsubsection{C$^{34}$S}

We derived the column density of C$^{34}$S and the H$_{2}$ density using 
the large velocity gradient (LVG) model \citep{gol74}. 
The method is the same as that adopted in \citet{hir98}. 
We ignored the difference in the beam sizes between the $J$=1-0 and 2-1 lines 
in the analysis. 
Because the core size of L492 (100\arcsec) is sufficiently larger than the 
beam sizes for the $J$=1-0 (36\arcsec) and $J$=2-1 line (20\arcsec), 
this approximation does not seriously affect the results \citep[e.g.][]{hir98}. 
The H$_{2}$ density is derived to be 9.0$\times$10$^{4}$ cm$^{-3}$.  
 
\subsubsection{CCS}

We derived the column density of CCS and the H$_{2}$ density using 
the LVG model. 
Because the beam size of the $J_{N}$=$2_{1}$-$1_{0}$ lines are as large as 
the core size, beam dilution effect would be significant, and hence, 
we only used the results of the 
$J_{N}$=$4_{3}$-$3_{2}$ and $J_{N}$=$7_{6}$-$6_{5}$ lines. 
For these lines, we ignored the difference in the beam sizes of 
the two transitions, as in the case of the C$^{34}$S analysis. 
The H$_{2}$ density is derived to be 
6.8$\times$10$^{4}$ cm$^{-3}$. 

For comparison, we carried out LTE analysis assuming the excitation temperature of 
5 K \citep{suz92, hir98, hir02, hir04}. In this case, column density of CCS is 
three times larger than that derived from the LVG analysis. Therefore, 
the column densities of CCS for other cores obtained by the LTE analysis, in which 
the excitation temperature is fixed to 5 K, would be overestimated. 
In this paper, the LTE value is employed for consistency with previous results. 

\subsubsection{DCO$^{+}$, DNC, H$^{13}$CO$^{+}$, HN$^{13}$C}

We derived the column densities of DCO$^{+}$, DNC, H$^{13}$CO$^{+}$, 
and HN$^{13}$C by the LVG calculations. The method is described in 
\citet{hir01}. Because we did not carry out multi-transition 
observations for these molecules, we assumed the H$_{2}$ density to be 
9.0$\times$10$^{4}$ cm$^{-3}$, 
which is derived from the C$^{34}$S data. 
If we assume the H$_{2}$ density to be 2.0$\times$10$^{5}$ cm$^{-3}$ 
\citep{cra05}, the column densities of DCO$^{+}$ and H$^{13}$CO$^{+}$ 
decrease by 10-25\%, while those of DNC and HN$^{13}$C 
decrease by a factor of 2. 
In both case, the DCO$^{+}$/H$^{13}$CO$^{+}$ 
and DNC/HN$^{13}$C ratios change by less than 10\%. 

\subsubsection{$^{13}$CO, C$^{18}$O}

The column densities of $^{13}$CO and C$^{18}$O were derived by assuming 
the $^{13}$CO/C$^{18}$O ratio of the local ISM value, 7.3 \citep{wil94}, 
and the common excitation temperature. 
The excitation temperature was determined to be 8.3 K. 

\subsubsection{HC$_{3}$N}

At first, we carried out the LVG analysis using the $J$=5-4 and 9-8 lines to 
derive the column density of HC$_{3}$N and the H$_{2}$ density. However, 
the fitting did not converge probably because of the large optical depth 
of the $J$=5-4 line. Because the satellite hyperfine component of the 
$J$=5-4 line is clearly detected as shown in Figure \ref{fig-sphc3n}, 
we can determine the optical depth and the excitation temperature of 
the $J$=5-4 line of HC$_{3}$N. The total optical depth and the excitation 
temperature were calculated to be 8.8 and 6.4 K, respectively.
The excitation temperature determined here is consistent with that employed 
by \citet{suz92}, 6.5 K. Then we determined the column density of 
HC$_{3}$N from the total optical depth assuming the LTE condition with 
the excitation temperature of 6.4 K. 

\subsubsection{C$_{3}$S, HC$_{5}$N, HC$_{7}$N}

For C$_{3}$S, HC$_{5}$N, and HC$_{7}$N, we determined their 
column densities by the consistent way adopted in the 
previous literatures \citep[e.g.][]{hir02, hir04, cer86}. 
We assumed that the excitation temperatures of 5.5 K for C$_{3}$S, 
6.5 K for HC$_{5}$N, and 10.0K for HC$_{7}$N. 

\subsubsection{N$_{2}$H$^{+}$}

Since all the hyperfine components of the N$_{2}$H$^{+}$ line are detected 
as shown in Figure \ref{fig-spn2hp}, we were able to determine 
the optical depth and the excitation temperature of N$_{2}$H$^{+}$ to be 
6.1 and 4.9 K, respectively. 
The excitation temperature determined here is consistent with those reported 
by \citet{cas02a}, 5 K. 
Then we derived the column density of N$_{2}$H$^{+}$ assuming the LTE condition 
with the excitation temperature of 4.9 K.  

\subsubsection{NH$_{3}$}

Using the intensity ratios of hyperfine components of the (1,1) lines, 
we derived the excitation temperature to be 4.1 K. 
In addition, we determined the rotation temperature of the para NH$_{3}$ 
to be 9.5${+1.7} \atop {-1.0}$ K by the methods described in \citet{ho83}. 
Thus, we calculated the column density of NH$_{3}$ assuming the LTE 
condition with the rotation temperature of 10 K and the ortho-to-para ratio of 1. 


\section{Discussions}

The column densities of the observed molecules are summarized in Table \ref{tab-column}. 
For comparison, the corresponding 
column densities in the carbon-chain--producing regions are 
also listed in Table \ref{tab-column}. 
The column densities of the carbon-chain molecules in L492 are significantly 
higher than those in typical dark cloud cores \citep{suz92} and 
are comparable to those in the carbon-chain--producing regions 
such as L1495B, L1521B, L1521E, and TMC-1. 
Since the carbon-chain--producing regions have so far been recognized in the Taurus 
Molecular Cloud, L492 is the first dark cloud core with extremely high abundances 
of carbon-chain-molecules that is found outside the Taurus Molecular Cloud. 
On the other hand, the column densities of NH$_{3}$ and N$_{2}$H$^{+}$ 
in L492 are higher than those in the carbon-chain--producing regions, 
and are comparable to those in typical dark cloud cores 
\citep{ben89, suz92, cas02a}. 
We stress that the column densities in L492 are 
almost comparable to those of TMC-1 for most of the observed molecules, 
except for the slightly high column density 
of NH$_{3}$ and the low column densities of sulfur-bearing carbon-chain molecules 
and HC$_{7}$N. Thus the chemical composition in L492 is close to 
that of TMC-1. 

In order to discuss chemical evolutionary stages of L492 and other starless cores, 
we summarize the selected molecular abundance ratios which are thought to be 
used as a chemical clock, as shown in Table \ref{tab-ratio}. 
The most popular indicator is the  NH$_{3}$/CCS ratio \citep{suz92}. 
The NH$_{3}$/CCS ratio of 6.5 in L492 is intermediate between 
those of carbon-chain--producing regions (2.6-3.8) and 
evolved prestellar core such as L1498 and L1544 (15-25). 
For consistency, we employed the NH$_{3}$/CCS ratios in L1498 and L1544 
reported by \citet{suz92}. 
Uncertainties in the NH$_{3}$/CCS ratios 
are quite large due to difference in telescope beam sizes, observed positions, 
and methods of data analysis. In fact, different values are reported by 
\citet{aik05} (14-25 and 9.0 for L1498 and L1544, respectively). 
If we use the result of the LVG calculations for the column density of CCS, 
the NH$_{3}$/CCS ratio in L492 is 21, 
which is larger by a factor of 3 than that of the LTE value. 
Even such uncertainties are considered, we can say that the NH$_{3}$/CCS ratio in 
L492 is intermediate or close to those of L1498 and L1544 rather 
than those in carbon-chain--producing regions. 

Another good indicator for chemical evolutionary stage is a deuterium 
fractionation ratio. Because deuterium fractionation proceeds
through the exothermic isotope exchange reaction of H$_{3}^{+}$ and HD in the gas-phase, 
the H$_{2}$D$^{+}$/H$_{3}^{+}$ ratio gradually increase with time 
\citep[e.g.][]{rob00, tur01, hir01, sai02}. 
In addition, depletion of molecules make the life time of 
H$_{2}$D$^{+}$ longer, resulting in a high deuterium fractionation ratio of 
H$_{2}$D$^{+}$/H$_{3}^{+}$. Consequently, 
a deuterium fractionation ratio is thought to be enhanced in evolved cores 
\citep[e.g.][]{rob00, tur01, hir01, sai02}. 
The deuterium fractionation ratio, DNC/HN$^{13}$C, of 1.27 in L492 
is comparable to those of the carbon-chain--producing regions (0.66-1.25), 
and is smaller than those in L1498 (1.91) and L1544 (3.0). 
The DCO$^{+}$/H$^{13}$CO$^{+}$ ratio of 
0.80 is also comparable to those of the carbon-chain--producing regions 
(0.63-1.10) and smaller than those in L1498 (2.7) and L1544 (3.1-9.2). 
The lower DCO$^{+}$/H$^{13}$CO$^{+}$ ratio in L492 suggests 
that molecular depletion is not very significant in comparison with 
L1544 \citep{cas02b}. 
The N$_{2}$D$^{+}$/N$_{2}$H$^{+}$ ratio of 0.05 in L492 is comparable to that in 
L1498 (0.04), while it is lower than in L1544 (0.23) \citep{cra05}. 
These results suggest that chemical evolutionary stage of L492 would be younger 
than those of L1498 and L1544, and molecules would be less depleted in L492 than 
in L1498 and L1544. 

\section{Evolutionary Scenario of L492 and Other Cores}

Recently, discussions considering both dynamical and chemical evolution 
of dense cores have been reported \citep{lee03, aik05, cra05, shi05}. 
For comparison, we summarize the evolutionary stages of L492 and other 
starless cores on the basis of signature of dynamical collapse, degree of depletion, and 
molecular abundance ratios, as shown in Table \ref{tab-scenario}. 
As already reported, L492 is in the dynamically collapsing phase 
as well as L1498 and L1544 \citep{oha99, lee01, cra05}, while 
L1521B, L1521E, and TMC-1 are not \citep{lee99, hir02, hir04}. 
Therefore, L492 can be regarded as a dynamically evolved core. 
\citet{cra05} recently reported that a degree of depletion of CO 
in L492 is comparable to that in 
L1498 and is lower than in L1544. On the other hand, the depletion of CO 
would not occur significantly in the 
carbon-chain--producing regions such as L1521E \citep{taf04}. 
Our results on the CS and CCS distributions suggest that depletion of 
these molecules seems to be 
less significant than in L1498 and L1544 \citep{taf02, taf04b}. 
Therefore, a degree of depletion of molecules 
in L492 is intermediate between chemically young cores (L1521E) and 
evolved cores (L1498 and L1544). 
The deuterium fractionation ratios of 
DNC/HN$^{13}$C and DCO$^{+}$/H$^{13}$CO$^{+}$ in L492 
are lower than those in L1498 and L1544,
and are comparable to those in carbon-chain--producing regions. This means 
that L492 is chemically younger than L1498 and L1544. Finally, 
the NH$_{3}$/CCS ratio in L492 is larger than those in carbon-chain--producing 
regions and is smaller than in L1498 and L1544, 
suggesting that L492 is more evolved than in carbon-chain--producing regions 
but is younger than L1498 and L1544. 

Considering these results, we conclude that the evolutionary stage of L492 is 
intermediate between the 
carbon-chain--producing regions (L1495B, L1521B, L1521E, and TMC-1) 
and evolved prestellar cores (L1498 and L1544). 
We propose the following scenario for chemical and dynamical evolution. 
At the initial state of dense cores before protostellar collapse, 
such as L1495B, L1521B, L1521E, and TMC-1, abundances of carbon-chain 
molecules are extremely high because of high abundance of the neutral carbon atom 
\citep{suz92}. On the other hand, NH$_{3}$ and N$_{2}$H$^{+}$ are not yet abundant 
because they are produced by slow chemical reactions \citep{suz92, ber97, aik05}. 
In addition, the deuterium fractionation in these cores are 
not enhanced because of the low H$_{2}$D$^{+}$/H$_{3}^{+}$ ratio at the earliest 
stage of chemical evolution, and also because of a low degree of depletion of CO, CS, CCS, 
and other molecules \citep{rob00, tur01, hir01, sai02}. As a core evolves, 
abundances of carbon-chain molecules are still high while 
those of NH$_{3}$ and N$_{2}$H$^{+}$ 
increase by the gas-phase reactions \citep{suz92, ber97}, and hence, 
the NH$_{3}$/CCS ratio increases with time. 
This would be the stage of L492. 
For L492, depletion of CS and CCS is less significant 
than L1498 and L1544, while the CO molecules start to deplete 
\citep{cra05}. 
Depletion of molecules would be just started 
in the central part of L492, if the double peak structure of the 
H$^{13}$CO$^{+}$ ($J$=1-0) and CCS ($7_{6}$-$6_{5}$) maps reflect real 
abundance variation. 
At the earliest phase of molecular depletion such as in L492, the deuterium 
fractionation is not significantly affected by the depletion. 
In this phase, protostellar collapse has already started \citep{lee01}. 
As a core further evolves, the abundances of 
carbon-chain molecules decrease due to gas-phase reactions and freeze-out of molecules 
onto grains while abundances of NH$_{3}$ and N$_{2}$H$^{+}$ increase significantly 
\citep{suz92, ber97}. This results in high NH$_{3}$/CCS ratio. 
The L1498 core would be just around this stage. 
The increase in the deuterium fractionation seems to be different 
from molecule to molecule. 
The N$_{2}$D$^{+}$/N$_{2}$H$^{+}$ ratio increases more slowly 
than the DCO$^{+}$/HCO$^{+}$ and DNC/HN$^{13}$C ratios. 
At the most evolved phase like L1544, depletion of molecule is significant, and 
the deuterium fractionation ratio increases drastically. 

It is also likely that timescales of 
gas-phase chemical evolution, freeze-out of molecules onto grains, and 
dynamical collapse are different from region to region 
(e.g. Taurus and Aquila Rift). 
For example, \citet{lee03} and \citet{shi05} 
classified three types of prestellar cores; 
chemically evolved and dynamically young cores (L1512, L1498), 
chemically evolved and dynamically evolved cores (L1544), 
and chemically young and dynamically evolved cores (L1689B). 
\citet{lee03} suggest that L1689B, which is in the Ophiuchus region, 
has the different environment from those of L1512 and L1544 in the 
Taurus region. It should be noted that L492 is also a chemically young and 
dynamically evolved core outside the Taurus region. 
In order to understand this regional difference, further systematic observational 
studies and theoretical modeling would be needed. 


\acknowledgements

We are grateful to the staff of Nobeyama Radio Observatory for their assistance 
in observations. TH thanks to the Inoue Foundation for Science 
(Research Aid of Inoue Foundation for Science) for the financial support. 
This study is partly supported by Grant-in-Aid from Ministry of 
Education, Science, Sports and Culture of Japan (14204013 and 15071201).

{}

\newpage

\newpage

\begin{deluxetable}{lrlcccc}
\tabletypesize{\scriptsize}
\tablenum{1}
\tablewidth{0pt}
\tablecaption{Gauss fit parameters for the observed lines \label{tab-observe}}
\tablehead{
\colhead{} & \colhead{Frequency} & \colhead{} & 
  \colhead{$T_{a}^{*}$} &   $v_{lsr}$ & 
  \colhead{$\Delta v$} &  \colhead{$T_{rms}$} \\
\colhead{Transition} & \colhead{(MHz)} & \colhead{Position} & 
  \colhead{(K)} & \colhead{(km s$^{-1}$)} &
  \colhead{(km s$^{-1}$)} &  \colhead{(K)} }
\startdata
$^{13}$CO($J$=1-0)\tablenotemark{a}
                             & 110201.353 & (20\arcsec,20\arcsec)  & 2.5(12) & 7.92(15) & 1.6(10) & 0.3 \\
C$^{18}$O($J$=1-0)\tablenotemark{a}
                             & 109782.173 & (20\arcsec,20\arcsec)  & 1.20(15) & 7.69(5) & 0.62(13) & 0.15 \\
DCO$^{+}$($J$=1-0)\tablenotemark{a}
                              & 72039.331 & (20\arcsec,20\arcsec)  & 0.83(17) & 7.70(5) & 0.49(13) & 0.13 \\
DNC($J$=1-0)\tablenotemark{a}
                              & 76305.717 & (20\arcsec,20\arcsec)  & 0.67(17) & 7.67(9) & 0.73(22) & 0.09 \\
H$^{13}$CO$^{+}$($J$=1-0)\tablenotemark{a}
                              & 86754.330 & (20\arcsec,20\arcsec)  & 0.83(10) & 7.83(4) & 0.61(10) & 0.08 \\
HN$^{13}$C($J$=1-0)\tablenotemark{a}
                              & 87090.859 & (20\arcsec,20\arcsec)  & 0.47(13) & 7.82(10) & 0.72(28) & 0.07 \\
C$^{34}$S($J$=1-0)\tablenotemark{a}
                              & 48206.946 & (40\arcsec,0\arcsec)   & 0.71(17) & 7.82(6) & 0.48(14) & 0.08 \\
C$^{34}$S($J$=2-1)\tablenotemark{a}
                              & 96412.961 & (40\arcsec,0\arcsec)   & 0.47(9) & 7.74(4) & 0.37(8) & 0.06 \\
CCS($J_{N}$=$2_{1}$-$1_{0}$)  & 22344.030 & (40\arcsec,0\arcsec)   & 0.69(6) & 7.78(3) & 0.80(8) & 0.03 \\
CCS($J_{N}$=$4_{3}$-$3_{2}$)\tablenotemark{a}
                    & 45379.033 & (40\arcsec,0\arcsec)   & 1.51(10) & 7.73(2) & 0.48(4) & 0.07 \\
CCS($J_{N}$=$7_{6}$-$6_{5}$)\tablenotemark{a}
                    & 81505.208 & (40\arcsec,0\arcsec)   & 0.87(20) & 7.71(5) & 0.42(12) & 0.09 \\
C$_{3}$S($J$=4-3)             & 23122.983 & (40\arcsec,0\arcsec)   & 0.41(6) & 7.77(4) & 0.59(11) & 0.04 \\
HC$_{3}$N($J$=5-4,$F$=5-5)
                      & 45488.834 & (40\arcsec,0\arcsec)   & 0.35(13) & 7.76(3) & 0.81(7) & 0.09 \\
HC$_{3}$N($J$=5-4,$F$=6-5/5-4/4-3)\tablenotemark{a,b}
                              & 45490.306 & (40\arcsec,0\arcsec)   & 2.35(16) & 7.76(3) & 0.81(7) & 0.09 \\
HC$_{3}$N($J$=5-4,$F$=4-4)    & 45492.106 & (40\arcsec,0\arcsec)   & 0.21(13) & 7.76(3) & 0.81(7) & 0.09 \\
HC$_{3}$N($J$=9-8)\tablenotemark{a}
                              & 81881.462 & (40\arcsec,0\arcsec)   & 1.70(9) & 7.73(1) & 0.46(3) & 0.09 \\
HC$_{5}$N($J$=9-8)            & 23963.901 & (40\arcsec,0\arcsec)   & 1.00(8) & 7.81(3) & 0.75(7) & 0.03 \\
HC$_{5}$N($J$=17-16)\tablenotemark{a}
                              & 45264.720 & (40\arcsec,0\arcsec)   & 1.23(12) & 7.76(2) & 0.49(6) & 0.06 \\
HC$_{7}$N($J$=21-20)          & 23687.898 & (0\arcsec,0\arcsec)    & 0.18(4) & 7.65(6) & 0.56(15) & 0.02 \\
N$_{2}$H$^{+}$($J,F_{1},F$=1,1,0-0,1,1)\tablenotemark{a}
                   & 93171.621 & (20\arcsec,20\arcsec) & 0.28(10) & 7.75(2) & 0.44(4) & 0.08 \\
N$_{2}$H$^{+}$($J,F_{1},F$=1,1,2-0,1,2)\tablenotemark{a}
                   & 93171.917 & (20\arcsec,20\arcsec) & 0.52(11) & 7.75(2) & 0.44(4) & 0.08 \\
N$_{2}$H$^{+}$($J,F_{1},F$=1,1,1-0,1,0)\tablenotemark{a}
                   & 93172.053 & (20\arcsec,20\arcsec) & 0.48(11) & 7.75(2) & 0.44(4) & 0.08 \\
N$_{2}$H$^{+}$($J,F_{1},F$=1,2,2-0,1,1)\tablenotemark{a}
                   & 93173.480 & (20\arcsec,20\arcsec) & 0.59(10) & 7.75(2) & 0.44(4) & 0.08 \\
N$_{2}$H$^{+}$($J,F_{1},F$=1,2,3-0,1,2)\tablenotemark{a}
                   & 93173.777 & (20\arcsec,20\arcsec) & 0.92(11) & 7.75(2) & 0.44(4) & 0.08 \\
N$_{2}$H$^{+}$($J,F_{1},F$=1,2,1-0,1,1)\tablenotemark{a}
                   & 93173.967 & (20\arcsec,20\arcsec) & 0.40(11) & 7.75(2) & 0.44(4) & 0.08 \\
N$_{2}$H$^{+}$($J,F_{1},F$=1,0,1-0,1,2)\tablenotemark{a}
                   & 93176.265 & (20\arcsec,20\arcsec) & 0.67(10) & 7.75(2) & 0.44(4) & 0.08 \\
NH$_{3}$($J, K$=1, 1; VH)\tablenotemark{a,c}
                              & 23692.955 & (0\arcsec,0\arcsec)    & 0.19(7) & 7.87(4) & 0.86(10) & 0.04 \\
NH$_{3}$($J, K$=1, 1; H)\tablenotemark{a,c}
                              & 23693.895 & (0\arcsec,0\arcsec)    & 0.31(7) & 7.87(4) & 0.86(10) & 0.04 \\
NH$_{3}$($J, K$=1, 1; M)\tablenotemark{a,c}
                              & 23694.496 & (0\arcsec,0\arcsec)    & 0.74(8) & 7.87(4) & 0.86(10) & 0.04 \\
NH$_{3}$($J, K$=1, 1; L)\tablenotemark{a,c}
                              & 23695.095 & (0\arcsec,0\arcsec)    & 0.33(7) & 7.87(4) & 0.86(10) & 0.04 \\
NH$_{3}$($J, K$=1, 1; VL)\tablenotemark{a,c}
                              & 23696.037 & (0\arcsec,0\arcsec)    & 0.24(7) & 7.87(4) & 0.86(10) & 0.04 \\
NH$_{3}$(2,2)                 & 23722.634 & (0\arcsec,0\arcsec)    & 0.06(2) & 7.71(11) & 0.63(27) & 0.02 \\
\enddata
\tablecomments{The numbers in parenthesis represent 
three times the standard deviation in the Gaussian fit in unit of the last significant digits.}
\tablenotetext{a}{Maps of the lines is shown in Figure \ref{fig-maps}. }
\tablenotetext{b}{Three hyperfine components are blended. }
\tablenotetext{c}{Hyperfine components are blended. Each component is labeled as adopted by 
\citet{ung80} and Figure \ref{fig-spnh3}. }
\end{deluxetable}

\begin{deluxetable}{llcclrcccl}
\tabletypesize{\scriptsize}
\tablenum{2}
\tablewidth{0pt}
\tablecaption{Results of LVG calculations \label{tab-lvg}}
\tablehead{
\colhead{} & \colhead{} &
  \colhead{} & \colhead{$\mu$\tablenotemark{b}} &  \colhead{} & \colhead{} & $T_{ex}$ & 
  \colhead{$n$(H$_{2}$)} &  \colhead{$N$} & \colhead{} \\
\colhead{Molecule} & \colhead{Transition} & 
  \colhead{$S_{ul}$\tablenotemark{a}} & \colhead{(Debye)} & \colhead{Position} & 
  \colhead{$\tau_{0}$} & \colhead{(K)} & \colhead{($\times 10^{5}$cm$^{-3}$)} & 
  \colhead{($\times 10^{13}$cm$^{-2}$)} & \colhead{Reference}}
\startdata
DCO$^{+}$         & $J$=1-0   & 1.00 & 4.07 & (20\arcsec,20\arcsec) & 0.47   & 7.4   & 0.90\tablenotemark{c} & 0.109 & 1,2 \\
DNC               & $J$=1-0   & 1.00 & 3.05 & (20\arcsec,20\arcsec) & 1.9   & 4.4   & 0.90\tablenotemark{c} & 0.56 & 3,4 \\
H$^{13}$CO$^{+}$  & $J$=1-0   & 1.00 & 4.07 & (20\arcsec,20\arcsec) & 0.82   & 5.9   & 0.90\tablenotemark{c} & 0.137 & 1,2 \\
HN$^{13}$C        & $J$=1-0   & 1.00 & 3.05 & (20\arcsec,20\arcsec) & 2.2   & 3.9   & 0.90\tablenotemark{c} & 0.44 & 3,4 \\
C$^{34}$S         & $J$=1-0   & 1.00 & 1.96 & (40\arcsec,0\arcsec)  & 0.20   & 8.6   & 0.90   & 0.38 & 5,6 \\
                  & $J$=2-1   & 2.00 & 1.96 & (40\arcsec,0\arcsec)  & 0.79   & 4.7   &        &      &     \\
CCS & $J_{N}$=$4_{3}$-$3_{2}$ & 3.97 & 2.81 & (40\arcsec,0\arcsec)  & 0.43   & 9.0   & 0.68   & 1.61 & 7,8,9 \\ 
    & $J_{N}$=$7_{6}$-$6_{5}$ & 6.97 & 2.81 & (40\arcsec,0\arcsec)  & 0.54   & 7.1   &        &      &       \\
\enddata
\tablenotetext{a}{Intrinsic line strength. }
\tablenotetext{b}{Dipole moment. }
\tablenotetext{c}{The H$_{2}$ density derived from the C$^{34}$S data is assumed.}
\tablerefs{1: \citet{hae79}; 2: \citet{mon85}; 
  3: \citet{bla76}; 4: \citet{gre74}; 
  5: \citet{win68}; 6: \citet{gre78}; 
  7: \citet{mur90}; 8: \citet{yam90}; 9: \citet{wol97}}
\end{deluxetable}

\clearpage
\begin{deluxetable}{llcclrccll}
\tabletypesize{\scriptsize}
\tablenum{3}
\tablewidth{0pt}
\tablecaption{Results of LTE calculations \label{tab-lte}}
\tablehead{
\colhead{} & \colhead{} & 
 \colhead{} & \colhead{$\mu$\tablenotemark{b}} &  
 \colhead{} & \colhead{} &$T_{ex}$ &  \colhead{$N$} & \colhead{} & \colhead{} \\
\colhead{Molecule} & \colhead{Transition} &
  \colhead{$S_{ul}$\tablenotemark{a}} & \colhead{(Debye)} & \colhead{Position} & 
  \colhead{$\tau_{0}$} & \colhead{(K)} & \colhead{($\times 10^{13}$cm$^{-2}$)} & \colhead{NOTE} & \colhead{Reference}}
\startdata
$^{13}$CO      & $J$=1-0    & 1.00 & 0.11 & (20\arcsec,20\arcsec)  & 4.0   & 8.3  & 1960     & $^{13}$CO/C$^{18}$O=7.3      & 1 \\
C$^{18}$O      & $J$=1-0    & 1.00 & 0.11 & (20\arcsec,20\arcsec)  & 0.64   & 8.3  &  310     & $^{13}$CO/C$^{18}$O=7.3      & 1 \\
CCS & $J_{N}$=$4_{3}$-$3_{2}$ & 3.97 & 2.81 & (40\arcsec,0\arcsec) & 3.4   & 5.0  &    5.3   & $T_{ex}$ fixed               & 2,3 \\
C$_{3}$S       & $J$=4-3    & 4.00 & 3.64 & (40\arcsec,0\arcsec)   & 0.21   & 5.5  &    0.55  & $T_{ex}$ fixed               & 2,3 \\
HC$_{3}$N      & $J$=5-4    & 5.00 & 3.72 & (40\arcsec,0\arcsec)   & 8.8   & 6.4  &   17.2   & hf fit\tablenotemark{c}      & 4 \\
HC$_{5}$N      & $J$=17-16 & 17.00 & 4.33 & (40\arcsec,0\arcsec)   & 0.65   & 6.5  &    4.1   & $T_{ex}$ fixed               & 5 \\
HC$_{7}$N      & $J$=21-20 & 21.00 & 4.82 & (0\arcsec,0\arcsec)    & 0.03  & 10.0  &    0.32  & $T_{ex}$ fixed               & 6 \\
N$_{2}$H$^{+}$ & $J$=1-0    & 1.00 & 3.40 & (20\arcsec,20\arcsec)  & 6.1   & 4.9  &    0.66  & hf fit\tablenotemark{c}      & 7 \\
NH$_{3}$       & $(1,1)$    & 1.50 & 1.46 & (0\arcsec,0\arcsec)    & 2.3   & 4.1  &   34     & hf fit\tablenotemark{c}, $T_{rot}$=10 K    & 8 \\
\enddata
\tablenotetext{a}{Intrinsic line strength. }
\tablenotetext{b}{Dipole moment. }
\tablenotetext{c}{Total optical depth and excitation temperatures are derived from a 
simultaneous fit to the hyperfine (hf) components. }
\tablerefs{1: \citet{goo94}; 2: \citet{mur90}; 3: \citet{yam90}; 4: \citet{laf78}; 5: \citet{ale76}; 
  6: \citet{bot97}; 7: \citet{hav90}; 8: \citet{coh74}}
\end{deluxetable}

\begin{deluxetable}{lccccccc}
\tabletypesize{\scriptsize}
\tablenum{4}
\tablewidth{0pt}
\tablecaption{Column densities of the selected molecules in unit of $10^{13}$ cm$^{-2}$
\label{tab-column}}
\tablehead{
\colhead{Molecule} & \colhead{L492}  &
  \colhead{L1495B} & \colhead{L1521B} &   
  \colhead{L1521E} & \colhead{TMC-1}  &    
  \colhead{Taurus} & \colhead{Others}
}
\startdata
C$^{34}$S        & 0.38   & \nodata                  & 0.56\tablenotemark{a}    & 1.25\tablenotemark{a}    & 0.73\tablenotemark{a}  &
  0.17-2.18\tablenotemark{a} & 0.08-0.68\tablenotemark{a}  \\
CCS              & 5.3\tablenotemark{b}  & 1.44\tablenotemark{c}    & 3.6\tablenotemark{d}     & 2.8\tablenotemark{c}     & 6.6\tablenotemark{d}   &
  $<$0.09-6.6\tablenotemark{d} & $<$1.1-1.51\tablenotemark{d} \\
C$_{3}$S         & 0.55   & 0.68\tablenotemark{e,f}  & 1.6\tablenotemark{e}     & 1.4\tablenotemark{g}     & 1.3\tablenotemark{d}   &
  0.42-1.3\tablenotemark{d}    & \nodata                     \\
HC$_{3}$N        & 17.2   & 2.1\tablenotemark{e}     & 4.1\tablenotemark{d}     & 2.3\tablenotemark{g}     & 17.1\tablenotemark{d}  &
  $<$0.07-17.1\tablenotemark{d} & $<$0.07-1.35\tablenotemark{d} \\
HC$_{5}$N        & 4.1    & 0.52\tablenotemark{e}    & 1.2\tablenotemark{e}     & 0.46\tablenotemark{g}    & 5.6\tablenotemark{d}   &
  $<$0.14-5.6\tablenotemark{d} & $<$0.17-1.07\tablenotemark{d} \\
HC$_{7}$N        & 0.32   & \nodata                  & 0.24\tablenotemark{h}    & \nodata                  & 1.4\tablenotemark{h}   &
  0.08-1.4\tablenotemark{h}    & \nodata \\
NH$_{3}$         & 34     & 5.5\tablenotemark{e}     & 12.6\tablenotemark{e}    & 7.3\tablenotemark{g}     & 19\tablenotemark{d}    &
  $<$2-107\tablenotemark{d} & $<$3-112\tablenotemark{d} \\
N$_{2}$H$^{+}$   & 0.66   & $<$0.17\tablenotemark{e} & $<$0.19\tablenotemark{e} & $<$0.14\tablenotemark{g} & 0.74\tablenotemark{i}  &
  0.3-1.3\tablenotemark{j}   & 0.14-2.7\tablenotemark{j}  \\
H$^{13}$CO$^{+}$ & 0.137  & 0.046\tablenotemark{c}   & 0.063\tablenotemark{c}   & 0.106\tablenotemark{c}   & 0.14\tablenotemark{k}  &
  0.038-0.164\tablenotemark{l}  & $<$0.0110-0.126\tablenotemark{l} \\
C$^{18}$O        & 310    & 190\tablenotemark{m}     & 300\tablenotemark{m}     & 170\tablenotemark{m}     & 330\tablenotemark{k}   &
  60-520\tablenotemark{l}  & 30-700\tablenotemark{l} \\
\enddata
\tablecomments{Positions for the cores are as follows: 
L492:   $\alpha_{2000}=18^{h}15^{m}46^{s}.1$, $\delta_{2000}=-03^{\circ}$46\arcmin13\arcsec \ and Tables \ref{tab-lvg} and \ref{tab-lte}; 
L1495B: $\alpha_{2000}=04^{h}15^{m}36^{s}.5$, $\delta_{2000}=+28^{\circ}$47\arcmin06\arcsec; 
L1521B: $\alpha_{2000}=04^{h}24^{m}12^{s}.7$, $\delta_{2000}=+26^{\circ}$36\arcmin53\arcsec; 
L1521E: $\alpha_{2000}=04^{h}29^{m}16^{s}.5$, $\delta_{2000}=+26^{\circ}$13\arcmin50\arcsec; 
TMC-1:  $\alpha_{2000}=04^{h}41^{m}42^{s}.5$, $\delta_{2000}=+25^{\circ}$41\arcmin27\arcsec. }
\tablenotetext{ a}{\citet{hir98}.}
\tablenotetext{ b}{The LTE value is employed here for consistency with other results. }
\tablenotetext{ c}{\citet{hir01}.}
\tablenotetext{ d}{\citet{suz92}.}
\tablenotetext{ e}{\citet{hir04}.}
\tablenotetext{ f}{Position is (80\arcsec,40\arcsec) offset from the reference position.}
\tablenotetext{ g}{\citet{hir02}.}
\tablenotetext{ h}{\citet{cer86}.}
\tablenotetext{ i}{\citet{hir95}.}
\tablenotetext{ j}{\citet{cas02a}.}
\tablenotetext{ k}{\citet{pra97}.}
\tablenotetext{ l}{\citet{but95}.}
\tablenotetext{ m}{\citet{mye83}.}
\end{deluxetable}

\begin{deluxetable}{lccccccc}
\tabletypesize{\scriptsize}
\tablenum{5}
\tablewidth{0pt}
\tablecaption{Molecular abundance ratios as an indicator of chemical evolution 
\label{tab-ratio}}
\tablehead{
\colhead{Molecule} & 
  \colhead{L1495B} &  \colhead{L1521B} &
  \colhead{L1521E} &  \colhead{TMC-1(CP)} & 
  \colhead{L492}   &  \colhead{L1498} & \colhead{L1544}  
}
\startdata
DNC/HN$^{13}$C             & $<$0.66\tablenotemark{a}   & 0.70\tablenotemark{a}   & 0.66\tablenotemark{b}
   & 1.25\tablenotemark{c} &  1.27 & 1.91\tablenotemark{a} & 3.0\tablenotemark{c}       \\
DCO$^{+}$/H$^{13}$CO$^{+}$ & 1.05\tablenotemark{a}      & 1.10\tablenotemark{a}   & 0.63\tablenotemark{a}
   & 0.77\tablenotemark{d} &  0.80 & 2.7\tablenotemark{e} & 3.1-9.2\tablenotemark{f}  \\
NH$_{3}$/CCS               & 3.8\tablenotemark{a,g}    & 3.5\tablenotemark{g,h} & 2.6\tablenotemark{a,b}
   & 2.9\tablenotemark{h} & 6.5 & 25\tablenotemark{h} & 15\tablenotemark{h,i} \\
\enddata 
\tablecomments{Positions for the cores are as follows: 
L1498: $\alpha_{2000}=04^{h}10^{m}51^{s}.5$, $\delta_{2000}=+25^{\circ}$09\arcmin58\arcsec; 
L1544: $\alpha_{2000}=05^{h}04^{m}16^{s}.6$, $\delta_{2000}=+25^{\circ}$10\arcmin48\arcsec. }
\tablenotetext{ a}{\citet{hir01}.}
\tablenotetext{ b}{\citet{hir02}.}
\tablenotetext{ c}{\citet{hir03}.}
\tablenotetext{ d}{\citet{tur01}.}
\tablenotetext{ e}{\citet{but95}.}
\tablenotetext{ f}{\citet{cas99}.}
\tablenotetext{ g}{\citet{hir04}.}
\tablenotetext{ h}{\citet{suz92}.}
\tablenotetext{ i}{Position is (-19\arcsec,60\arcsec) offset from the reference position.}
\end{deluxetable}

\begin{deluxetable}{lccccccc}
\tabletypesize{\scriptsize}
\tablenum{6}
\tablewidth{0pt}
\tablecaption{Evolutionary scenario of dark cloud cores
\label{tab-scenario}}
\tablehead{
\colhead{Item} & 
  \colhead{L1495B} &  \colhead{L1521B} &
  \colhead{L1521E} &  \colhead{TMC-1(CP)} & 
  \colhead{L492}   &  \colhead{L1498} & \colhead{L1544}  
}
\startdata
Gas-phase molecular abundance  &    &    &    &    &    &    &   \\
\hline
DNC/HN$^{13}$C\tablenotemark{a}
                               & Y  & Y  & Y  & Y  & Y  & E  & E \\ 
DCO$^{+}$/H$^{13}$CO$^{+}$\tablenotemark{a}
                               & Y  & Y  & Y  & Y  & Y  & E  & E \\ 
N$_{2}$D$^{+}$/N$_{2}$H$^{+}$\tablenotemark{b}
                               & ? & ? & Y  & Y  & E  & E  & E \\
NH$_{3}$/CCS\tablenotemark{a}
                               & Y  & Y  & Y  & Y  & I  & E  & E \\ 
\hline
Degree of depletion            &    &    &    &    &    &    &   \\
\hline
Depletion factor of CO\tablenotemark{b}
                               & ? & ? & Y  & ? & I  & I  & E \\ 
Map structure of CS and CCS\tablenotemark{c}
                               & Y  & Y  & Y  & Y  & I?  & E  & E \\ 
\hline
Kinematics                     &    &    &    &    &    &    &   \\
\hline
Signature of infall\tablenotemark{d}
                               & ?  & Y  & Y  & Y  & E  & E  & E \\ 
\enddata 
\tablecomments{Y, I, and E indicates young, intermediate, and evolved, respectively. }
\tablenotetext{ a}{Table \ref{tab-ratio}}
\tablenotetext{ b}{\citet{cra05}}
\tablenotetext{ c}{\citet{hir02,hir04,taf02,taf04}}
\tablenotetext{ d}{\citet{lee99,lee01,cra05}}
\end{deluxetable}

\end{document}